\documentclass[
superscriptaddress,amsmath,amssymb,aps,longbibliography,twocolumn,showpacs,a4paper]{revtex4-2}
\usepackage{graphicx}
\usepackage{mathtools}
\usepackage{bm}
\usepackage{braket}
\usepackage{xcolor}
\usepackage[utf8]{inputenc}
\usepackage{booktabs}
\usepackage{siunitx}
\usepackage[margin=1in]{geometry}
\usepackage[version=4]{mhchem}
\usepackage[colorlinks=true,linkcolor=blue,citecolor=blue,urlcolor=blue]{hyperref}

\newcommand{\LMSO}{\ce{La2O3Mn2Se2}}

\usepackage[normalem]{ulem} 

\makeatletter

\renewcommand{\section}{\@startsection{section}{1}{\z@}
  {-2.5ex \@plus -1ex \@minus -.2ex}
  {1.0ex \@plus .2ex}
  {\normalfont\large\bfseries\raggedright}}

\renewcommand{\subsection}{\@startsection{subsection}{2}{\z@}
  {-2.25ex\@plus -1ex \@minus -.2ex}
  {0.8ex \@plus .2ex}
  {\normalfont\normalsize\bfseries\raggedright}}

\def\@seccntformat#1{\csname the#1\endcsname\hskip 0.5em}
\makeatother
\date{\today}

\begin{document}
\title{Pressure-Tunable Electronic and Magnonic Transport in Altermagnet \LMSO}

\author{Nafise Rezaei}
\affiliation{Skolkovo Institute of Science and Technology, 121205, Bolshoy Boulevard 30, bld. 1, Moscow, Russia.}
\author{Alireza Qaiumzadeh}
\email{alireza.qaiumzadeh@ntnu.no}
\affiliation{Center for Quantum Spintronics, Department of Physics, Norwegian University of Science and Technology, NO-7491 Trondheim, Norway}
\author{Artem R. Oganov}
\affiliation{Skolkovo Institute of Science and Technology, 121205, Bolshoy Boulevard 30, bld. 1, Moscow, Russia.}
\affiliation{Sber University, Universitetskaya St. 11, Anosino Village, Istra District, 143581, Russia}
\author{Mojtaba Alaei}
\email{m.alaei@skoltech.ru}
\affiliation{Skolkovo Institute of Science and Technology, 121205, Bolshoy Boulevard 30, bld. 1, Moscow, Russia.}
\affiliation{Department of Physics, Isfahan University of Technology, Isfahan 84156-83111, Iran.}

\begin{abstract}
Hydrostatic pressure provides a symmetry-preserving route to engineer electronic and magnonic transport in the correlated insulating altermagnet \LMSO. Using first-principles calculations combined with spin-Hamiltonian modeling, we show that compression from 0 to 40 GPa markedly enhances the inequivalence between the competing second-neighbor exchange interactions, increasing $|J_{2a}-J_{2b}|$ from 1.97 to 9.38 meV while preserving the compensated antiferromagnetic ground state. The resulting exchange anisotropy amplifies the momentum-dependent splitting between the two chiral magnon branches, yielding a nearly fourfold enhancement of the longitudinal magnon-driven spin Seebeck response at 100 K, from $3.68\times10^{-1}$ to $1.36$ meV/K.
 In contrast, hydrostatic pressure preserves the magnetic-symmetry selection rules governing the anomalous Hall effect while redistributing the electronic Berry curvature, producing pronounced energy-dependent sign reversals in the anomalous Hall conductivity. These results identify exchange anisotropy as the microscopic mechanism underlying the pressure-enhanced magnon response and establish hydrostatic pressure as an effective means of simultaneously controlling electronic and magnonic transport in insulating altermagnets.
\end{abstract}

\maketitle
\section{Introduction}
Altermagnets are a class of nonrelativistic $\mathcal{PT}$-broken collinear antiferromagnets in which opposite-spin sublattices are related by crystal rotations or mirror operations rather than by translation or inversion. This unique spin-group symmetry gives rise to symmetry-protected momentum-dependent spin splitting despite zero net magnetization~\cite{Smejkal2022PRX,Smejkal2022Landscape}. This symmetry enables spin-polarized electronic states despite a vanishing net magnetization, giving rise to transport phenomena commonly associated with ferromagnets, including spin-polarized currents and anomalous Hall effects~\cite{GonzalezHernandez2021,Smejkal2020SciAdv}. It also constrains the spin Hamiltonian, allowing chiral spin-polarized magnon bands~\cite{Smejkal2023ChiralMagnons} and establishing altermagnets as a promising platform for electronic and magnonic transport \cite{weissenhofer2026magnon,PhysRevB.111.134448, krk8-655j, zwz9-l7wf}.

Although the electronic and magnonic manifestations of altermagnetism have attracted considerable attention, they have largely been investigated separately. Electronic studies have primarily focused on momentum-dependent spin splitting, Berry-curvature-driven transport, and anomalous Hall effects, predominantly in metallic systems~\cite{Smejkal2022Landscape,GonzalezHernandez2021,Smejkal2020SciAdv}, whereas magnonic studies have emphasized chiral spin-wave excitations and spin caloritronic phenomena in insulating magnets~\cite{Smejkal2023ChiralMagnons,Bauer2012,Chumak2015}. Consequently, practical strategies for simultaneously tuning Berry-curvature responses and magnon transport within a single insulating altermagnetic material remain largely unexplored.
Hydrostatic pressure is particularly attractive for this purpose because it continuously modifies the lattice geometry and magnetic exchange interactions without introducing chemical disorder \cite{PhysRevB.111.104416}. It therefore provides a symmetry-preserving route to engineer magnon transport through exchange reconstruction while simultaneously reshaping the electronic structure and the associated Berry-curvature distribution.

La$_2$O$_3$Mn$_2$Se$_2$ provides an ideal platform for realizing this strategy. It crystallizes in the tetragonal $I4/mmm$ structure and has recently been established as a correlated insulating $d$-wave altermagnet with compensated G-type antiferromagnetic order, in which the Mn sublattices are related by fourfold rotational symmetry and the momentum-dependent spin splitting follows the inverse-Lieb-lattice mechanism~\cite{Ni2010,Koo2012,Lei2012,Wei2024LMSO,Asai2026LMSO}. Its magnetic interactions arise from a delicate competition between direct exchange and multiorbital superexchange, making the exchange pathways highly sensitive to lattice compression \cite{GarciaGassull2025LMSO}. Moreover, recent high-pressure experiments have revealed substantial structural evolution, including an unusual in-plane lattice collapse, while preserving the crystal structure up to approximately 53~GPa~\cite{Li2025PressureLMSO}. Together, these characteristics make La$_2$O$_3$Mn$_2$Se$_2$ an excellent candidate for pressure engineering of both magnon transport and electronic Berry-curvature responses.

In this work, we demonstrate that hydrostatic pressure provides a symmetry-preserving route to simultaneously engineer magnon transport and electronic Berry-curvature responses in the insulating altermagnet \LMSO. Focusing on its compensated G-type magnetic ground state, we show that compression from 0 to 40~GPa markedly enhances the inequivalence between the competing direction-dependent second-neighbor isotropic exchange interactions, leading to increased chiral magnon splitting and a nearly fourfold enhancement of the longitudinal magnon-driven spin Seebeck response in the insulating phase. In contrast, the magnetic symmetry and the associated selection rules governing the anomalous Hall response remain unchanged under pressure, while the electronic Berry-curvature distribution is substantially redistributed, giving rise to pronounced energy-dependent sign reversals in the anomalous Hall conductivity (AHC) in the conducting doped system. These findings identify exchange reconstruction as the microscopic mechanism underlying the enhanced magnon transport while revealing a distinct pressure-driven evolution of Berry-curvature responses within the same insulating altermagnetic material.

\begin{figure}[t]
    \centering
    \includegraphics[width=1.\linewidth]{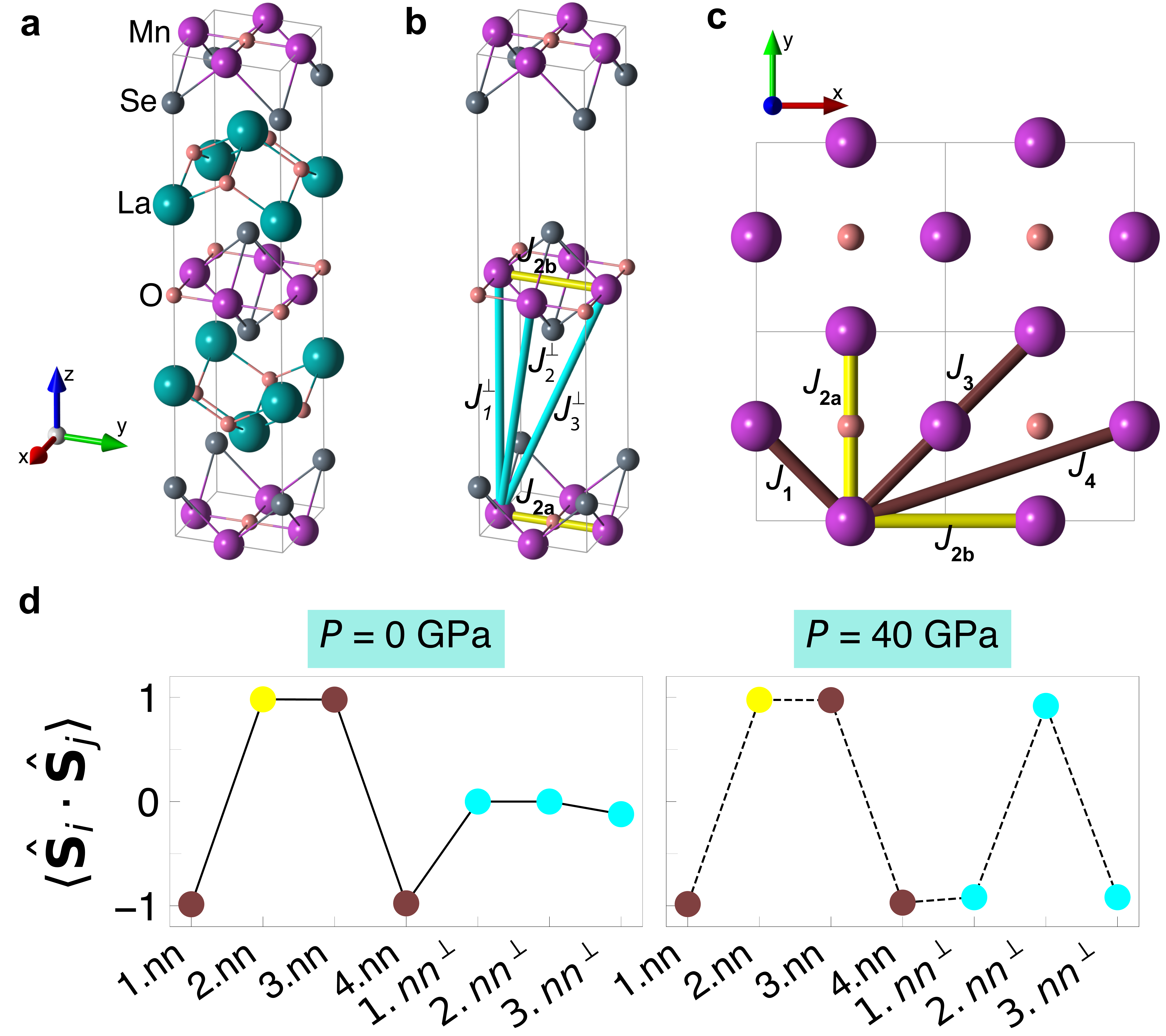}
    \caption{Crystal and magnetic-exchange geometry of \LMSO. (a) Tetragonal layered structure. (b,c) Intralayer and interlayer exchange pathways used in the spin Hamiltonian, highlighting the linear Mn--O--Mn second-neighbor bridge ($J_{2a}$) and the buckled Mn--Se--Mn second-neighbor bridge ($J_{2b}$). (d) Bond-resolved spin correlations from Monte Carlo simulations at $T=10$~K. Pressure leaves the intralayer antiferromagnetic pattern nearly unchanged but strengthens alternating G-type interlayer correlations at $40$~GPa.}
    \label{fig:struct}
\end{figure}

\begin{table*}[htbp]
\centering
\caption{Spin-Hamiltonian parameters and Monte Carlo-predicted Néel temperatures for \LMSO. The intralayer exchange interactions \(J_{1}\)--\(J_{4}\), interlayer exchange interactions \(J_{i}^{\perp}\), and biquadratic interaction \(B\), DM interaction parameter \(D\) are given in meV, while the Néel temperature \(T_{\mathrm{N}}\) is given in Kelvin. Values are listed for the experimental structure, the relaxed structure at \(0\)~GPa, and \(40\)~GPa.}
\label{tab:J}
\begin{tabular}{lccccccccccc}
\toprule
Structure & $J_1$ & $J_{2a}$ & $J_{2b}$ & $J_3$ & $J_4$ & $J_{1}^{\perp}$ & $J_{2}^{\perp}$ & $J_{3}^{\perp}$ & $B$ & $D$ & $T_\mathrm{N}$  \\ 
\midrule
Exp. & -32.06 & -3.87  & -6.49  & -0.79 & -0.23 & 0.014 & -0.013 & -0.016 & -2.67 & -0.20 & 202\\
$P=0$~GPa & -28.79 & -2.68  & -4.65  & -0.56 & -0.21 & 0.011 & -0.010 & -0.013 & -2.75 & -0.17& 198\\
$P=40$~GPa & -48.23 & -12.90 & -22.28 & -2.66 & -0.95 & 0.060 & -0.080 & -0.106 & -2.37 & -0.55 & 161\\
experiment\cite{Asai2026LMSO}  & -29.25 & -5.13 & -7.00 & & & & & & & & \\ 
calculation\cite{Asai2026LMSO} & -32.25 & -5.25 & -7.25 & & & & & & & & \\
\bottomrule
\end{tabular}
\end{table*}

\section{Results}
\begin{figure*}
        \centering
   \includegraphics[width=1.\linewidth,keepaspectratio]{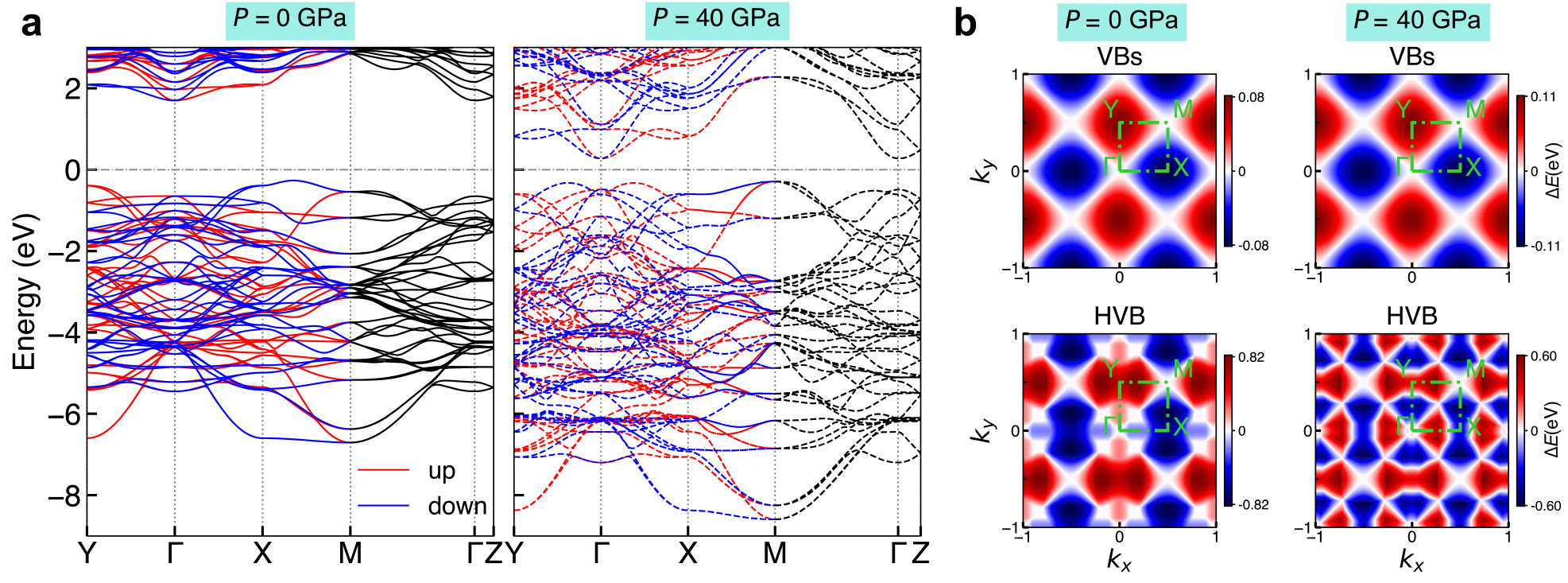}
   \caption{Pressure evolution of the altermagnetic electronic structure. (a) Spin-resolved band structures at $0$ and $40$~GPa. (b) Momentum-resolved spin splitting $\Delta E(\mathbf{k})$ on the $k_z=0$ plane, averaged over all valence bands (VBs) and shown separately for the highest valence band (HVB). Compression preserves the four-lobed $d$-wave spin texture while changing the magnitude of the altermagnetic splitting.}
    \label{fig:bands_vs_P}
\end{figure*}

\subsection{Electronic band structure}
Figure~\ref{fig:struct}(a) shows the crystal structure of \LMSO, which crystallizes in the centrosymmetric tetragonal space group $I4/mmm$ (No.~139).
The electronic structure at ambient pressure and 40~GPa is shown in Fig.~\ref{fig:bands_vs_P}. In both cases, the spin-resolved bands exhibit a characteristic $d$-wave momentum-dependent spin splitting. We define the average spin splitting
$\Delta E(\mathbf{k})
= N^{-1}\sum_{n=1}^{N}\left[E_n^{\uparrow}(\mathbf{k})-E_n^{\downarrow}(\mathbf{k})\right]$, where N is the the number of electronic bands,
evaluated for (i) all entangled valence bands (excluding semi-cores) and (ii) the highest valence band. 

In both cases, $\Delta E(\mathbf{k})$ exhibits a robust four-lobed $d$-wave pattern on the $k_z=0$ plane, dictated by the crystal symmetry and insensitive to the number of bands included. Under pressure, this symmetry is preserved while the bandwidth increases and the magnitude of spin splitting is enhanced, indicating a strengthened altermagnetic response without modification of its symmetry character.

\subsection{Spin Hamiltonian model}
We compute spin-spin interactions using the total energy mapping method \cite{rezaei2026benchmarking, Mosleh2023}. 
Figures~\ref{fig:struct}(b) and (c) show the dominant intra- and interlayer magnetic exchange pathways. The layered crystal structure and lattice symmetry give rise to two crystallographically distinct second-nearest-neighbor (2.nn) isotropic intralayer superexchange couplings: the linear Mn--O--Mn pathway ($J_{2a}$) and the buckled Mn--Se--Mn pathway ($J_{2b}$). The magnetic interactions are described by the bilinear Heisenberg Hamiltonian
\begin{equation}
\mathcal{H}_{J}
=
-\sum_{\substack{i>j\\{\rm intra}}} J_{ij}\,
\hat{\mathbf S}_i\cdot\hat{\mathbf S}_j
-\sum_{\substack{i>j\\{\rm inter}}} J_{ij}^{\perp}\,
\hat{\mathbf S}_i\cdot\hat{\mathbf S}_j .
\end{equation} 
where $J_{ij}$ and $J^{\perp}_{ij}$ denote intralayer and interlayer exchange interactions, respectively, and $\hat{\mathbf S}_i$ is a unit vector along the magnetic moment at site $i$. In this toy model, only the Mn ions carry magnetic moments and are therefore included as magnetic sites in the spin Hamiltonian. The calculated exchange parameters are summarized in Table~\ref{tab:J}. The magnetic energy scale is dominated by the strong intralayer antiferromagnetic 1.nn exchange interaction $J_1<0$, which stabilizes the Néel order within each Mn$_2$OSe$_2$ layer. In contrast, the interlayer exchange interactions are about three orders of magnitude weaker, establishing the quasi-2D magnetic character of \LMSO. Consequently, the 3D magnetic ground state is determined by the subtle competition among the weak interlayer exchange interactions.

The two crystallographically distinct 2.nn exchange pathways, $J_{2a}$ and $J_{2b}$, connect Mn atoms separated by the same Mn--Mn distance owing to the tetragonal symmetry. Consequently, their different pressure evolution cannot be inferred from the Mn--Mn separation alone. Indeed, the Mn--Mn distance decreases identically for both pathways, from $4.1666$~\AA\ at 0~GPa to $3.9864$~\AA\ at 40~GPa. The ligand geometry, however, changes differently for the two bridges. For $J_{2a}$, the Mn--O--Mn path remains linear, with an angle of $180^\circ$ at both pressures, and the Mn--O distance decreases from $2.0833$ to $1.9932$~\AA. For $J_{2b}$, the Mn--Se distance decreases more strongly, from $2.8382$ to $2.5919$~\AA, while the Mn--Se--Mn angle opens from $94.45^\circ$ to $100.53^\circ$. Pressure therefore nearly self-similarly compresses the linear oxygen-mediated $J_{2a}$ path, whereas it both shortens and opens the buckled selenium-mediated $J_{2b}$ path.
This distinct structural response of the two ligand-mediated exchange pathways is directly reflected in the calculated exchange interactions. Although both 2.nn couplings become increasingly antiferromagnetic under compression, the selenium-mediated interaction $J_{2b}$ strengthens substantially more than the oxygen-mediated interaction $J_{2a}$. Consequently, the inequivalence between the two competing 2.nn exchange interactions, quantified by $|J_{2a}-J_{2b}|$, increases from $1.97$~meV at ambient pressure to $9.38$~meV at $40$~GPa, providing the microscopic origin of the enhanced chiral magnon splitting discussed below. 

The interlayer exchange interactions are two to three orders of magnitude weaker than their intralayer counterparts and therefore require a dedicated high-accuracy determination. To this end, we evaluate 12 distinct magnetic configurations within a $2\times2\times1$ supercell. The magnetic configurations are constructed in pairs with identical intralayer spin arrangements but different interlayer stackings. Consequently, subtracting the corresponding total-energy equations eliminates the intralayer exchange contributions, yielding a system of equations that depends exclusively on the interlayer exchange parameters $J_{1}^{\perp}$, $J_{2}^{\perp}$, and $J_{3}^{\perp}$. The complete set of magnetic configurations is provided in the Supplementary Information (SI). The extracted interlayer exchange parameters are summarized in Table~\ref{tab:J}. 
The leading interlayer couplings, $J_{1}^{\perp}>0$ and $J_{3}^{\perp}<0$ (Fig.~\ref{fig:struct}(b)), favor different magnetic stacking between adjacent layers. Specifically, $J_{1}^{\perp}$ stabilizes C-type order (ferromagnetic interlayer stacking), whereas $J_{3}^{\perp}$ stabilizes G-type order (antiferromagnetic interlayer stacking), leading to competition between the two magnetic states.

Under hydrostatic compression, the signs of all exchange interactions remain unchanged, while their magnitudes increase (Table.~\ref{tab:J}). The interlayer couplings reach only $0.06$--$0.11$~meV at 40~GPa, remaining two to three orders of magnitude smaller than the dominant intralayer interactions. Thus, pressure reinforces the existing interlayer coupling without altering the quasi-2D magnetic character of \LMSO.

We further find that the G-type and C-type stacking configurations are nearly degenerate at ambient pressure, with energy differences comparable to the intrinsic accuracy of first-principles calculations. This near degeneracy persists across the PBE+$U$, SCAN, and HSE functionals, indicating that DFT does not establish a clear energetic preference for either interlayer stacking. The robust intralayer magnetic order and the associated exchange hierarchy, however, remain unaffected.


Although we found that G-type and C-type orders are almost degenerate at ambient pressure within our simulation accuracy, magnetization studies of \LMSO reported G-type ordering with a weak ferromagnetic behavior, initially associated with spin reorientation or canting \cite{Ni2010}. Ref. ~\cite{Wei2024LMSO} established, from the reported G-type $k=(0,0,0)$ order, a compensated d-wave altermagnetic state. Although a small net magnetic moment appears below about $140$ K, its origin remains unresolved, and magnetic neutron pair distribution function (mPDF) refinements show no spin canting at $100$ K. We therefore test whether higher-order or antisymmetric exchange terms can perturb the compensated collinear state.

To this end, we extend the bilinear exchange Hamiltonian $\mathcal{H}_{J}$ as
\begin{equation}
\mathcal{H}
=
\mathcal{H}_{J}
+ B \sum_{i>j}
\left(\hat{\mathbf{S}}_i \cdot \hat{\mathbf{S}}_j\right)^2
+ \sum_{i>j}
\mathbf{D}_{ij} \cdot
\left(
\hat{\mathbf{S}}_i \times \hat{\mathbf{S}}_j
\right),
\label{eq:spinH}
\end{equation}
where \(B\) is the nearest-neighbor biquadratic exchange and \(\mathbf{D}_{ij}\) is the Dzyaloshinskii--Moriya (DM) vector, both evaluated using the four-state method~\cite{4S-0,4S-1,4S-2}. The large negative \(B\) (Table~\ref{tab:J}) favors collinear spin configurations, with a slightly reduced magnitude under pressure. Symmetry allows a nearest-neighbor DM vector only along the \(z\) axis, giving an estimated canting angle \(\delta \simeq D/(J_1+2B)\). Fully noncollinear calculations yield only \(\delta \simeq 0.15^\circ\) at ambient pressure and no resolved canting at \(P=40\)~GPa within the angular resolution of
 \(0.025^\circ\), i.e., \(\delta < 0.025^\circ\). Thus, consistent with the mPDF results of Ref.~\cite{Wei2024LMSO}, the experimentally reported ferromagnetic component is not reproduced as an intrinsic net moment in our spin-Hamiltonian mapping.

We also calculated the magnetic anisotropy arising from spin--orbit coupling (SOC) and magnetic dipole--dipole interactions. For the SOC contribution, VASP yields magnetic anisotropy energies of 1 and 36~$\mu$eV per Mn atom at $P=0$ and $P=40$ GPa, respectively, both favoring an easy-plane magnetic anisotropy. Independent calculations using the FLEUR ~\cite{fleurCode, fleurWeb} code give corresponding values of 40 and 59~$\mu$eV per Mn atom, again favoring the easy plane. Although the quantitative values differ between the two implementations, the overall energy scale is very small, and we therefore expect the SOC anisotropy to be sensitive to details of the \textit{ab initio} methodology, such as the treatment of the DFT+$U$ correction.

In contrast to the SOC contribution, the magnetic dipole--dipole interaction favors an easy axis along the $z$ direction. The corresponding anisotropy energies are 87 and 90~$\mu$eV per Mn atom at $P=0$ and $P=40$ GPa, respectively. Consequently, the easy-axis anisotropy observed experimentally at ambient pressure is consistent with the dominance of the dipole--dipole interaction over the SOC contribution. At elevated pressure, however, the enhanced SOC anisotropy predicted by the \textit{ab initio} calculations becomes comparable to the dipolar contribution, suggesting that the preferred magnetization direction may change from easy-axis to easy-plane.

\subsection{Chiral Magnon Spectrum}
Using the exchange parameters obtained from first-principles calculations, we compute the magnon spectrum within linear spin-wave theory for localized spins with $S=5/2$. The calculations include the intralayer Heisenberg exchange interactions and the biquadratic exchange term. The interlayer exchange couplings are neglected because they are two to three orders of magnitude smaller than the intralayer interactions and primarily introduce weak magnetic frustration, with a negligible effect on the spin-wave dispersion. The DM interaction is also omitted, as its calculated strength is much smaller than the dominant exchange interactions.

\begin{figure}[htbp]
    \centering
    \includegraphics[width=0.5\textwidth]{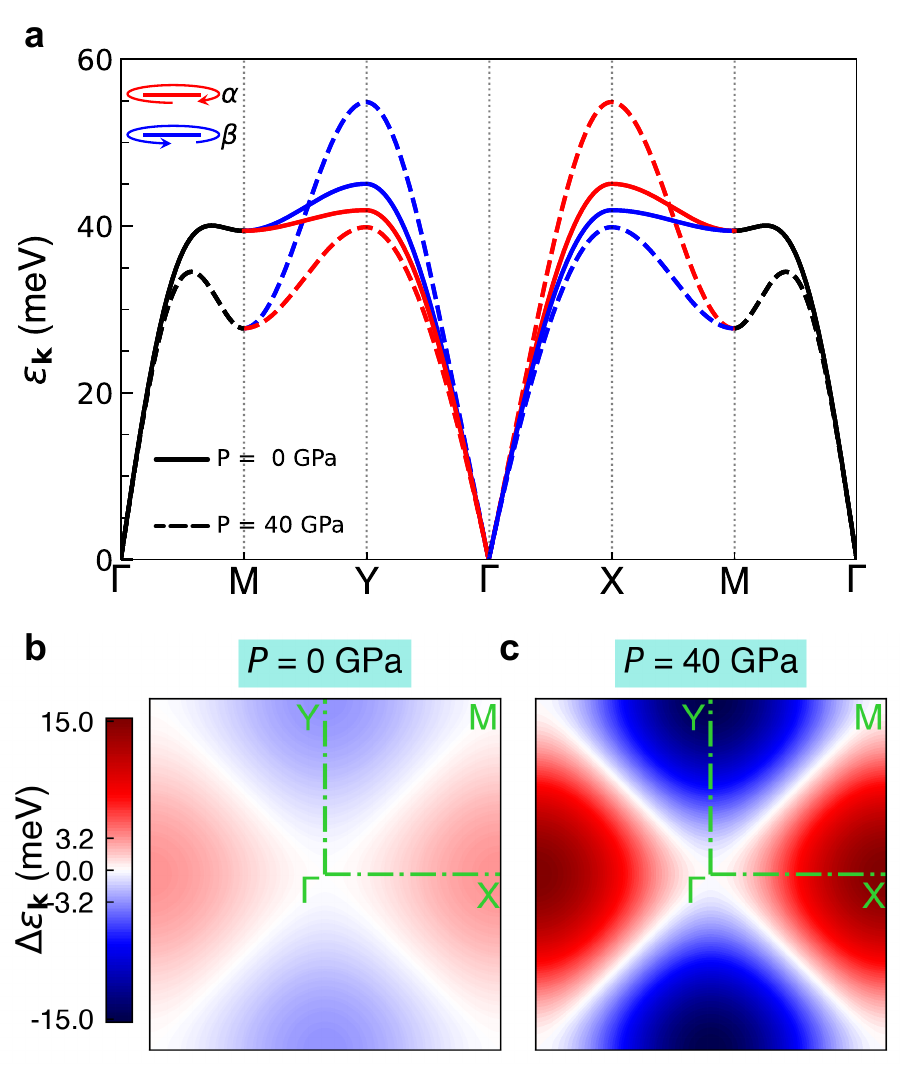}
    \caption{
Pressure-enhanced chiral magnon splitting. (a) Magnon energy $\varepsilon_{\mathbf{k}}$ (meV) along the high-symmetry path at $0$~GPa (solid lines) and $40$~GPa (dashed lines) pressures, with the $\alpha$ and $\beta$ branches shown in red and blue. (b,c) Branch splitting $\Delta\varepsilon_{\mathbf{k}}=\varepsilon_{\alpha\mathbf{k}}-\varepsilon_{\beta\mathbf{k}}$ (meV) in the $k_z=0$ plane at $0$ and $40$~GPa pressures. 
}
    \label{fig:magnon}
\end{figure}

Figure~\ref{fig:magnon} shows the noninteracting magnon dispersion at zero temperature under ambient pressure (solid lines) and at 40~GPa (dashed lines). The spectrum consists of two chiral magnon branches, denoted $\alpha$ (red) and $\beta$ (blue). In the altermagnetic state, the inequivalence between the 2.nn exchange interactions,
$\delta J_2 = |J_{2a}-J_{2b}|$,
lifts the degeneracy of the two branches along symmetry directions of the magnetic BZ, while the two branches remain degenerate at the $\Gamma$ point.

Hydrostatic pressure enhances $\delta J_2$, leading to a larger magnon-band splitting throughout the magnetic BZ. The maximum splitting occurs at the symmetry-related $X$ and $Y$ points, where the sign of the chiral splitting is reversed by symmetry and its magnitude follows the approximate scaling relation
$\Delta\varepsilon(X)\approx1.60\,\delta J_2$.
The momentum-resolved magnon splitting,
$\Delta\varepsilon_{\mathbf{k}}
=
\varepsilon_{\alpha\mathbf{k}}
-
\varepsilon_{\beta\mathbf{k}}$,
evaluated in the $k_z=0$ plane, Figs.~\ref{fig:magnon}(b) and (c), further illustrate this evolution.

\subsection{Monte Carlo prediction of finite-temperature magnetic order}
The finite-temperature magnetic properties of \LMSO~were investigated using classical Monte Carlo simulations. The magnetic specific heat $C_M(T)$ exhibits pronounced peaks at $T_{\mathrm N}\approx198$~K under ambient pressure and $T_{\mathrm N}\approx161$~K at 40~GPa. Although the dominant antiferromagnetic 1.nn exchange $J_1$ is enhanced under compression, the antiferromagnetic 2.nn couplings $J_{2a}$ and $J_{2b}$ increase even more strongly relative to $J_1$. These enhanced competing 2.nn interactions introduce stronger frustration of the intralayer N\'eel order, providing a natural explanation for the reduced estimated $T_{\mathrm N}$ at 40~GPa.
Experimentally, the transition to 3D long-range antiferromagnetic ordering occurs at a N\'eel temperature of 163 K, corresponding to a sharp cusp in zero-field-cooled susceptibility and a subtle heat-capacity anomaly~\cite{Ni2010}.

To characterize the magnetic ordering, we compute the bond-resolved spin-spin correlation functions shown in Fig.~\ref{fig:struct}(d). The temperature dependence of the magnetic specific heat $C_M(T)$ and staggered magnetization $M_{\rm stag}(T)$ is provided in the SI \cite{SM}. The latter is defined as the norm of the sublattice-weighted total spin,
$M_{\rm stag}={N_{\rm mag}^{-1}}\left|\sum_{i=1}^{N} c_i \hat{\mathbf{S}}_i\right|$,
where $N_{\rm mag}$ is the number of magnetic atoms (Mn) in the unit cell. $c_i=\pm1$ are phase factors chosen according to the G-type antiferromagnetic ordering pattern. The staggered magnetization quantifies the degree of long-range antiferromagnetic order, with $M_{\rm stag}=1$ corresponding to a perfectly ordered collinear state. 

At $T=10$~K, the intralayer spin correlations for the first four intralayer neighbor shells are
$\langle \hat{\mathbf{S}}_i \cdot \hat{\mathbf{S}}_j \rangle \approx (-1,\,+1,\,+1,\,-1)$
at both pressures (Fig.~\ref{fig:struct}(d)), demonstrating that the intralayer N\'eel order within each layer is robust against hydrostatic compression.

Across independent Monte Carlo simulations at $T=10$~K, and under ambient pressure, the staggered magnetization varies between $M_{\rm stag}\approx0.1$ and $0.5$, indicating strong fluctuations and the absence of complete 3D phase locking. This behavior is consistent with the near degeneracy of the competing G- and C-type stacking configurations, which originates from the competition between the leading interlayer exchange interactions, $J_{1}^{\perp}>0$ and $J_{3}^{\perp}<0$, favoring G-type and C-type interlayer stacking, respectively, as predicted by our first-principles calculations.

Under hydrostatic pressure, the interlayer correlations at $T=10$~K develop a clear alternating pattern characteristic of G-type stacking,
$
\langle \mathbf S_i \cdot \mathbf S_j \rangle = (-1,\,+1,\,-1)$,
for successive interlayer neighbor shells. At the same temperature, the staggered magnetization increases to $M_{\rm stag}\approx 0.96$, indicating substantially stronger 3D phase coherence within finite-size accuracy. These results show that pressure favors G-type interlayer coherence and drives a crossover away from quasi-2D magnetism, consistent with the pressure-enhanced interlayer exchange interactions.

\begin{figure}[t]
    \centering
    \includegraphics[width=0.5\textwidth,keepaspectratio]{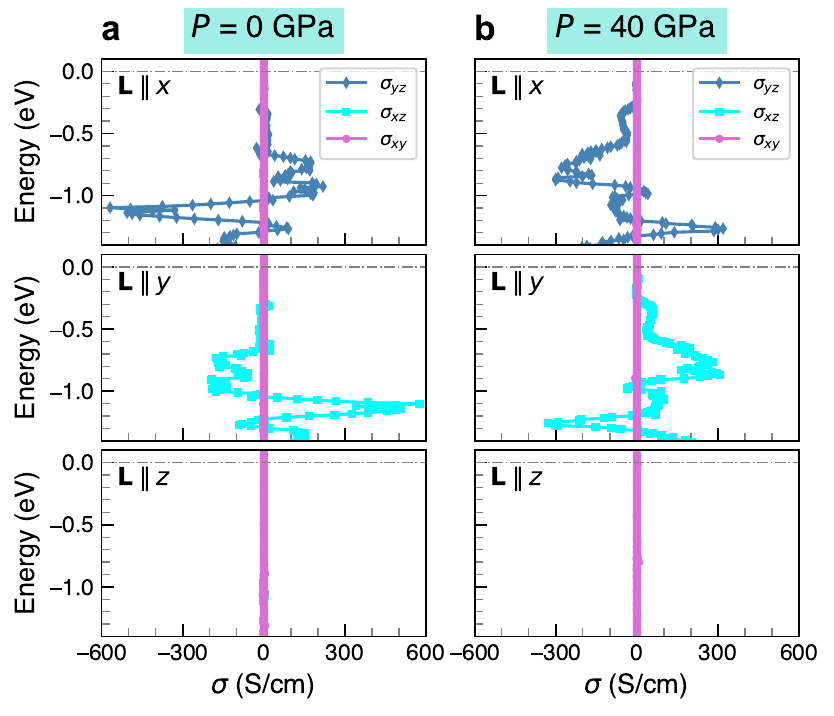}
    \caption{
Energy-resolved intrinsic AHC at (a) $0$ and (b) $40$~GPa for $\mathbf{L}\parallel x$, $\mathbf{L}\parallel y$, and $\mathbf{L}\parallel z$. The curves show $\sigma_{yz}$, $\sigma_{xz}$, and $\sigma_{xy}$ as functions of the scanned Fermi level, in S~cm$^{-1}$. The dashed horizontal line marks zero Hall conductivity; pressure preserves the symmetry-allowed tensor components while redistributing the Berry-curvature response over energy.
}
    \label{fig:ahc}
\end{figure}

\section{Electronic and Magnonic Transport}
We next investigate the transport properties of \LMSO. We first examine the Berry-curvature-driven electronic transport as a function of the Fermi energy, followed by the magnon-mediated longitudinal spin Seebeck transport in the insulating state.

\subsection{Anomalous Hall Conductivity}
First, we evaluate the intrinsic AHC to probe the pressure evolution of the Berry-curvature response. The AHC is calculated from the Berry curvature of the spin-orbit-coupled Wannier-interpolated bands according to
\[
\sigma_{ij}=-\frac{e^2}{\hbar}
\sum_n\int_{\rm BZ}\frac{d^3 \mathbf{k}}{(2\pi)^3}
f_n(\mathbf{k})\Omega_{n}^{ij}(\mathbf{k}),
\]
where $i \neq j \in \{x,y,z\}$, $f_n(\mathbf{k})$ is the Fermi-Dirac distribution evaluated at the chosen Fermi level, and $\Omega_{n}^{ij}(\mathbf{k})$ is the Berry curvature of band $n$. Figure~\ref{fig:ahc} shows the calculated AHC as the Fermi level is scanned across the electronic band structure and for different magnetic ground states.

The magnetic point-group symmetry determines which components of the AHC tensor survive the Brillouin-zone integration. For the G-type magnetic structure, only $\sigma_{yz}$ is allowed for the N\'eel vector oriented along the $x$-axis ($\mathbf{L}\parallel x$), while only $\sigma_{xz}$ is allowed for $\mathbf{L}\parallel y$. In contrast, all AHC tensor components are symmetry forbidden for $\mathbf{L}\parallel z$. As shown in Fig.~\ref{fig:ahc}, these symmetry selection rules remain unchanged under hydrostatic pressure: the forbidden tensor components remain negligibly small throughout the scanned energy window, whereas the entire AHC response originates exclusively from the symmetry-allowed components.

At ambient pressure, the symmetry-allowed AHC components, $\sigma_{yz}$ for $\mathbf{L}\parallel x$ and $\sigma_{xz}$ for $\mathbf{L}\parallel y$, reach peak magnitudes of about $6\times10^2$~S~cm$^{-1}$, reflecting a strongly energy-dependent Berry-curvature imbalance accumulated below the scanned chemical potential. At $40$~GPa, the peak magnitudes of these symmetry-allowed components are reduced to about $3\times10^2$~S~cm$^{-1}$, and their spectra develop positive and negative lobes of comparable size across the Fermi-level scan. This sign-changing behavior indicates a pressure-induced redistribution of the Berry-curvature contributions to the AHC, while the magnetic-symmetry selection rules remain unchanged.

\begin{figure}[tb!]
    \centering
    \includegraphics[width=0.5\textwidth]{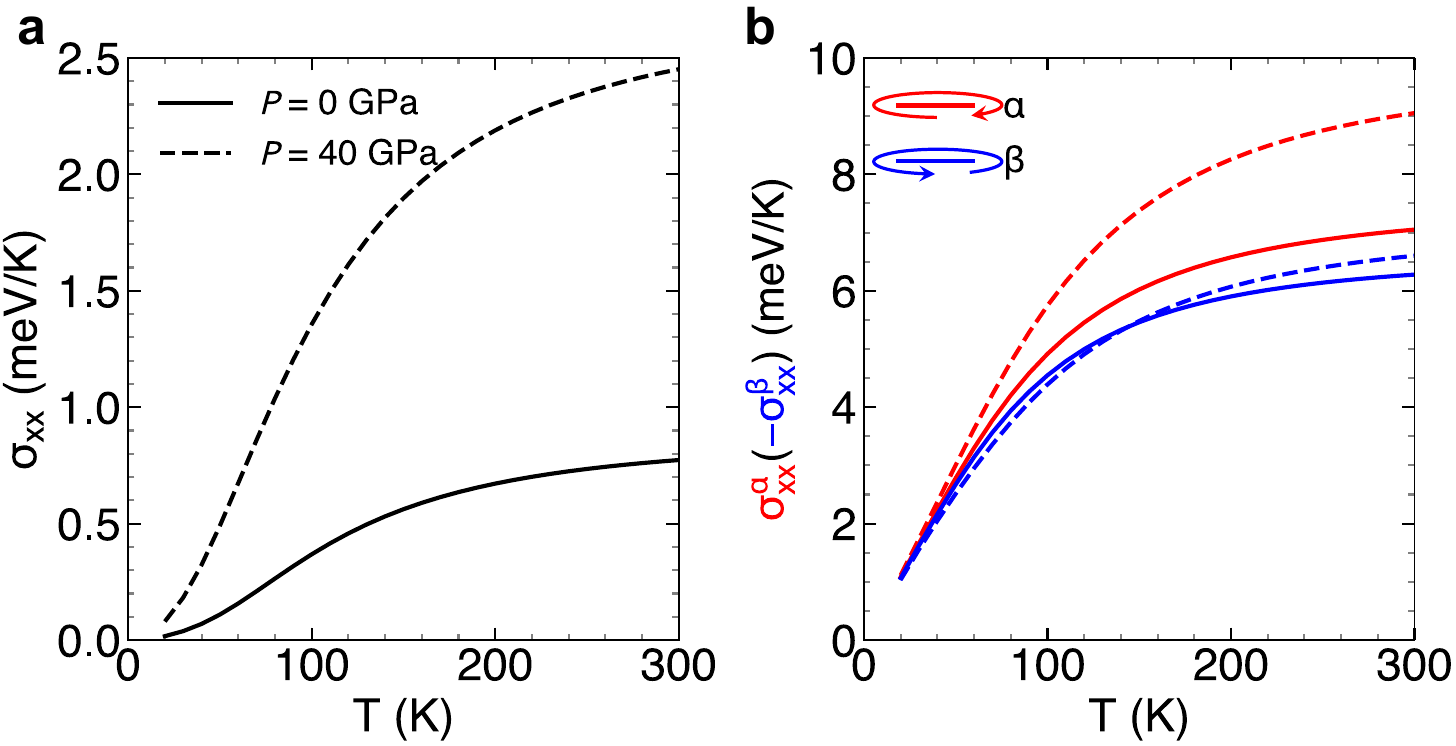}
    \caption{
Spin-Seebeck response from unequal $\alpha$- and $\beta$-magnon transport. (a) Total longitudinal spin Seebeck conductivity $\sigma_{xx}$ as a function of temperature at $0$ and $40$~GPa. (b) Branch-resolved magnitudes $\sigma_{xx}^{\alpha}$ and $-\sigma_{xx}^{\beta}$, plotted with the same pressure line styles as in (a). All conductivities are given in meV/K. Pressure enhances the total response by increasing the imbalance between opposite-spin magnon branches, mainly through the $\alpha$ channel.
}
    \label{fig:SSE}
\end{figure}

\subsection{Spin Seebeck Response}
In $\mathcal{PT}$-symmetric easy-axis collinear antiferromagnets, the two chiral magnon branches are degenerate in the absence of an external magnetic field Consequently, thermally excited magnons with opposite spin polarization contribute equally to spin transport, yielding no net magnon-driven spin Seebeck response. In contrast, the altermagnetic exchange anisotropy lifts this degeneracy, producing an energy imbalance between the two chiral magnon branches. As shown in Fig.~\ref{fig:magnon}, hydrostatic pressure further enhances this momentum-dependent splitting, thereby increasing the thermal population imbalance between magnons with opposite spin polarization and enhancing the magnon-driven spin Seebeck response.

The spin Seebeck effect is described by the linear response relation
$\mathbf{J}_s=-\boldsymbol{\sigma}\nabla T$,
where $\mathbf{J}_s$ is the magnon-mediated spin current induced by the temperature gradient $\bm{\nabla} T$, and $\boldsymbol{\sigma}$ is the longitudinal spin Seebeck conductivity. Within the Boltzmann transport formalism for altermagnetic magnons~\cite{Cui2023EfficientSSE}, the magnon conductivity of the $\nu=\alpha,\beta$ band is given by
{\small
\begin{align}
\sigma^\nu_{mn}
&=
-\frac{\tau_0}{\hbar AN_{\mathbf{k}}k_BT^2}
\sum_{\mathbf{k}}
S_{\nu\mathbf{k}}^{z}
V_{m,\nu\mathbf{k}}V_{n,\nu\mathbf{k}}
\varepsilon_{\nu\mathbf{k}} \nonumber
\\
&\quad\times
\frac{e^{\varepsilon_{\nu\mathbf{k}}/(k_BT)}}
{\left[e^{\varepsilon_{\nu\mathbf{k}}/(k_BT)}-1\right]^2},
\end{align}}
and the total magnon conductivity is $\sigma_{mn}=\sigma^\alpha_{mn}+\sigma^\beta_{mn}$, where the indices $m$ and $n$ denote the Cartesian directions ($x$ and $y$).
Here, $\tau_0 = 10$ ps is the assumed magnon relaxation time, chosen as a typical value (see Ref.~\cite{Cui2023EfficientSSE}), $A$ is the system area, and the conductivity is computed on a magnetic Brillouin-zone mesh of $N_k$ points.
The velocity factor is evaluated from the band-energy derivative, $V_{m,\nu\mathbf{k}}=\partial_{k_m}\varepsilon_{\nu\mathbf{k}}$. The branch spin angular momentum is evaluated from the field derivative of the magnon energy,
$S^{z}_{\nu\mathbf{k}}=
-(g\mu_B)^{-1}{\partial \varepsilon_{\nu\mathbf{k}}}/{\partial B_z}$.

The crystal symmetry imposes the constraint
$\sigma_{xx}^{\alpha}= -\sigma_{yy}^{\beta}$,$\sigma_{yy}^{\alpha}= -\sigma_{xx}^{\beta}$,
and $\sigma_{xy}^{\alpha}=\sigma_{xy}^{\beta}=0$, 
which follows from the combined lattice rotation and sublattice-resolved spin inversion symmetry.

As shown in Fig.~\ref{fig:SSE}, hydrostatic pressure enhances the longitudinal magnon-driven spin Seebeck conductivity at $T=100$~K from
$\sigma_{xx}(P=0)=3.68\times10^{-1}$ to
$\sigma_{xx}(P=40)=1.36$~(meV/K).
This enhancement originates from the pressure-induced increase in the momentum-resolved splitting of the two chiral magnon branches [Fig.~\ref{fig:magnon}], which amplifies the thermal population imbalance between magnons with opposite spin polarization.

\section{Discussion}
We have demonstrated that hydrostatic pressure provides a symmetry-preserving route for simultaneously controlling electronic and magnonic transport in the insulating altermagnet \LMSO. Although the magnetic symmetry and the associated selection rules remain unchanged under compression, pressure drives a pronounced reconstruction of both the electronic and magnetic excitation spectra through its effect on the underlying exchange interactions.

Our calculations reveal two distinct consequences of this exchange reconstruction. On the magnonic side, the enhanced inequivalence between the competing second-neighbor exchange interactions increases the momentum-dependent splitting of the two chiral magnon branches, leading to a substantial enhancement of the longitudinal magnon-driven spin Seebeck response. On the electronic side, pressure redistributes the Berry curvature without altering the symmetry-allowed structure of the anomalous Hall effect, resulting in pronounced energy-dependent changes in the AHC.

An attractive implication of these results is that similar modifications of the exchange interactions may also be achieved through chemical pressure. For example, partial or complete substitution of La by smaller nonmagnetic cations, such as Y, Sc, Lu, or Yb, would reduce the lattice volume through ionic-size effects while preserving the absence of magnetic moments on the rare-earth site. Likewise, partial substitution of Se by the smaller S atom provides an additional route to lattice contraction. Such chemically induced structural tuning may reproduce the pressure-driven enhancement of the exchange anisotropy and the associated magnonic and electronic transport responses, offering a practical alternative to externally applied hydrostatic pressure.

These results establish exchange anisotropy as the microscopic link between lattice compression and magnon transport in insulating altermagnets, while demonstrating that hydrostatic pressure provides a general strategy for engineering magnonic and Berry-curvature responses within the same material platform. It would be interesting to explore whether similar effects can also be achieved through chemical pressure, for example, by substituting La with smaller nonmagnetic ions (e.g., Y, Sc, Lu, or Yb) or Se with S.

\section{Methods}
\subsection{First-principles calculations}
Most density functional theory (DFT) calculations were performed in the Vienna Ab initio Simulation Package (VASP)~\cite{Kresse1993VASP,Kresse1996VASP} using the projector-augmented-wave method~\cite{Blochl1994PAW,Kresse1999PAW}. Exchange and correlation were treated with PBE~\cite{Perdew1996PBE} plus a Dudarev Hubbard correction~\cite{Dudarev1998DFTU}. Following previous first-principles studies of \LMSO~\cite{Koo2012,Asai2026LMSO}, an effective Hubbard parameter of $U_{\mathrm{eff}}=4$ eV was applied to the Mn ($3d$) states. Production calculations used a 550 eV plane-wave cutoff and a $\Gamma$-centered k-point mesh with reciprocal-space resolution of approximately 0.25~$\mathrm{\AA}^{-1}$.

The 0 and 40 GPa structures were obtained by relaxing lattice vectors and internal coordinates under hydrostatic pressure until the forces, the accuracy of the stress components, and electronic self-consistency were below
0.0015 eV $\mathrm{\AA}^{-1}$, 0.01 GPa, and \(10^{-8}\) eV, respectively

Exchange constants were fitted to collinear total energies. Four intralayer interactions were extracted from more than 60 magnetic configurations in a 63-atom SUPERHEX supercell~\cite{Alaei2024SUPERHEX}. Three interlayer interactions were fitted from nine configurations in a \(2\times2\times1\), 72-atom supercell.

The biquadratic and DM interactions were calculated using the four-state method in a $2 \times 2 \times 1$ supercell containing 36 atoms, generated from the Niggli-reduced unit cell.

To validate the magnetic anisotropy energies obtained with VASP, we carried out independent calculations using the FLEUR code, an all-electron DFT package based on the full-potential linearized augmented plane-wave (FLAPW) method. Because FLEUR is well suited for noncollinear magnetic calculations, we also employed it to determine the equilibrium canting angle.

All crystal structures of \LMSO\ used in this study, including the experimental structure~\cite{Wei2024LMSO} and the DFT+$U$-optimized structures, are provided in the SI~\cite{SM}.
\subsection{Monte Carlo simulations}
Classical Monte Carlo simulations used ESpinS~\cite{Rezaei2019ESpinS} on a \(14\times14\times3\) spin cell. Each temperature used \(10^{6}\) equilibration steps and \(10^{6}\) averaging steps, sampled every 4 steps. Parallel tempering exchanges were attempted every 20 steps~\cite{Hukushima1996ParallelTempering}.

\subsection{Magnon dispersion and spin Seebeck calculations}
Magnon spectra were calculated in SpinW~\cite{Toth2015SpinW} from the fitted exchange parameters. 

Spin Seebeck conductivities were computed from the branch-resolved Boltzmann expression using \(\tau_0=10\) ps and a \(300\times300\times66\) mesh. The spin angular momentum \(S_{\nu\mathbf{k}}^z\) was obtained by finite differences of \(\varepsilon_{\nu\mathbf{k}}=\hbar\omega_{\nu\mathbf{k}}\) under \(B_z=10^{-4}\) T.

\subsection{Wannier interpolation and anomalous Hall conductivity}
The maximally localized Wannier functions were constructed from density-functional calculations using a reciprocal-space sampling defined by a \(k\)-point spacing of \(0.15~\text{\AA}^{-1}\). Averaged spin splitting \(\Delta E(\mathbf{k})\) maps were obtained from Wannier interpolation without spin–orbit coupling on a dense \(100\times100\times100\) \(k\)-mesh. AHC and Berry-curvature energy maps were calculated from separate spin–orbit--coupled Wannier Hamiltonians generated with Wannier90~\cite{Mostofi2008Wannier90,Marzari1997MLWF,Souza2001MLWF}. The basis included Mn \(s,d\), Se \(p\), La \(d,f\), and O \(p\) orbitals. Conductivities were evaluated on a \(50\times50\times50\) coarse mesh with twofold adaptive refinement~\cite{Wang2006AHC}. 

\section{Data availability}
The numerical data supporting the findings of this study, including relaxed structural models, exchange parameters, spin-correlation data, magnon spectra, spin-Seebeck conductivities, spin-resolved band data, and AHC spectra, are available from the corresponding author upon reasonable request. 

\section{Author contributions}
N.R. performed the majority of the calculations, including DFT, MC simulations, electron, and magnon transport calculations. M.A. carried out several additional DFT calculations and contributed to the analysis. The project was conceived by N.R., M.A., and A.Q., A.R.O. provided guidance and support throughout the computational work. The initial manuscript was written by N.R. and subsequently revised and edited by A.Q., M.A., and A.R.O.

\section{Acknowledgments}
A.Q. was supported by the Research Council of Norway through Grant Nos. 353919 and 361800 ``QTransMag'', and Grant No. 262633 ``QuSpin''.

\bibliography{references}

@article{Smejkal2022PRX,
  author = {Smejkal, Libor and Sinova, Jairo and Jungwirth, Tomas},
  title = {Beyond Conventional Ferromagnetism and Antiferromagnetism: A Phase with Nonrelativistic Spin and Crystal Rotation Symmetry},
  journal = {Phys. Rev. X},
  volume = {12},
  pages = {031042},
  year = {2022},
  doi = {10.1103/PhysRevX.12.031042}
}

@article{PhysRevB.111.134448,
  title = {Efficient generation of spin currents in altermagnets via magnon drag},
  author = {Sourounis, Konstantinos and Manchon, Aur\'elien},
  journal = {Phys. Rev. B},
  volume = {111},
  issue = {13},
  pages = {134448},
  numpages = {9},
  year = {2025},
  month = {Apr},
  publisher = {American Physical Society},
  doi = {10.1103/PhysRevB.111.134448},
  url = {https://link.aps.org/doi/10.1103/PhysRevB.111.134448}
}

@article{PhysRevB.111.104416,
  title = {Origin of $A$-type antiferromagnetism and chiral split magnons in altermagnetic $\ensuremath{\alpha}$-{MnTe}},
  author = {Alaei, Mojtaba and Sobieszczyk, Pawel and Ptok, Andrzej and Rezaei, Nafise and Oganov, Artem R. and Qaiumzadeh, Alireza},
  journal = {Phys. Rev. B},
  volume = {111},
  issue = {10},
  pages = {104416},
  numpages = {7},
  year = {2025},
  month = {Mar},
  publisher = {American Physical Society},
  doi = {10.1103/PhysRevB.111.104416},
  url = {https://link.aps.org/doi/10.1103/PhysRevB.111.104416}
}

@article{rezaei2026benchmarking,
  title={Benchmarking first-principles approaches for extracting magnetic exchange interactions},
  author={Rezaei, Nafise and Oganov, Artem R and Ghojavand, Ali and Milo{\v{s}}evi{\'c}, Milorad V and Alaei, Mojtaba},
  journal={npj Computational Materials},
  year={2026},
  publisher={Nature Publishing Group},
  doi={https://doi.org/10.1038/s41524-026-02161-3},
  url={https://doi.org/10.1038/s41524-026-02161-3}
}

@article{Mosleh2023,
  title = {Benchmarking density functional theory on the prediction of antiferromagnetic transition temperatures},
  author = {Mosleh, Zahra and Alaei, Mojtaba},
  journal = {Phys. Rev. B},
  volume = {108},
  issue = {14},
  pages = {144413},
  numpages = {12},
  year = {2023},
  month = {Oct},
  publisher = {American Physical Society},
  doi = {10.1103/PhysRevB.108.144413},
  url = {https://link.aps.org/doi/10.1103/PhysRevB.108.144413}
}

@article{4S-0,
  title = "{Predicting the spin-lattice order of frustrated systems from first principles}",
  author = {Xiang, H. J. and Kan, E. J. and Wei, Su-Huai and Whangbo, M.-H. and Gong, X. G.},
  journal = {Phys. Rev. B},
  volume = {84},
  issue = {22},
  pages = {224429},
  numpages = {5},
  year = {2011},
  month = {Dec},
  publisher = {American Physical Society},
  doi = {10.1103/PhysRevB.84.224429},
  url = {https://link.aps.org/doi/10.1103/PhysRevB.84.224429}
}

@article{4S-1,
    author = {Xiang, Hongjun and Lee, Changhoon and Koo, Hyun-Joo and Gong, Xingao and Whangbo, Myung-Hwan},
    title = {Magnetic properties and energy-mapping analysis},
    journal = {Dalton Transactions},
    volume = {42},
    number = {4},
    pages = {823-853},
    year = {2013},
    month = {01},
    issn = {1477-9226},
    doi = {10.1039/c2dt31662e},
    url = {https://doi.org/10.1039/c2dt31662e},
    eprint = {https://pubs.rsc.org/dt/article-pdf/42/4/823/2659394/c2dt31662e.pdf},
}

@Article{4S-2,
AUTHOR = {Li, Xueyang and Yu, Hongyu and Lou, Feng and Feng, Junsheng and Whangbo, Myung-Hwan and Xiang, Hongjun},
TITLE = {Spin Hamiltonians in Magnets: Theories and Computations},
JOURNAL = {Molecules},
VOLUME = {26},
YEAR = {2021},
NUMBER = {4},
ARTICLE-NUMBER = {803},
URL = {https://www.mdpi.com/1420-3049/26/4/803},
PubMedID = {33557181},
ISSN = {1420-3049},
DOI = {10.3390/molecules26040803}
}

@MISC{fleurWeb,
  author = {},
  title = {{The FLEUR project}},
  howpublished = {\url{https://www.flapw.de/}}
}

@misc{fleurCode,
  author       = {Wortmann, Daniel and Michalicek, Gregor and Baadji, Nadjib and Betzinger, Markus and Bihlmayer, Gustav and Br\"oder, Jens and Burnus, Tobias and Enkovaara, Jussi and Freimuth, Frank and Friedrich, Christoph and Gerhorst, Christian-Roman and Granberg Cauchi, Sabastian and Grytsiuk, Uliana and Hanke, Andrea and Hanke, Jan-Philipp and Heide, Marcus and Heinze, Stefan and Hilgers, Robin and Janssen, Henning and Kl\"uppelberg, Daniel Aaaron and Kovacik, Roman and Kurz, Philipp and Lezaic, Marjana and Madsen, Georg K. H. and Mokrousov, Yuriy and Neukirchen, Alexander and Redies, Matthias and Rost, Stefan and Schlipf, Martin and Schindlmayr, Arno and Winkelmann, Miriam and Bl\"ugel, Stefan},
  title        = {{FLEUR}},
  month        = may,
  year         = 2023,
  publisher    = {Zenodo},
  doi          = {10.5281/zenodo.7576163},
  url          = {https://doi.org/10.5281/zenodo.7576163},
  howpublished  = {Zenodo}
}

@article{weissenhofer2026magnon,
  title={Magnon orbital Nernst effect in altermagnets},
  author={Wei{\ss}enhofer, Markus and Mrudul, MS and Mankovsky, Sergiy and Oppeneer, Peter M},
  journal={npj Quantum Mater.},
  year={2026},
  url={https://doi.org/10.1038/s41535-026-00853-z},
  publisher={Nature Publishing Group UK London}
}

@article{krk8-655j,
  title = {Quantum geometry and magnon Hall transport in an altermagnet},
  author = {Sylju\aa{}sen, Erlend and Qaiumzadeh, Alireza and Sudb\o{}, Asle},
  journal = {Phys. Rev. B},
  volume = {112},
  issue = {6},
  pages = {064429},
  numpages = {9},
  year = {2025},
  month = {Aug},
  publisher = {American Physical Society},
  doi = {10.1103/krk8-655j},
  url = {https://link.aps.org/doi/10.1103/krk8-655j}
}

@article{zwz9-l7wf,
  title = {Phonon-enhanced optical spin conductivity and spin-splitter effect in altermagnets},
  author = {Hodt, Erik Wegner and Qaiumzadeh, Alireza and Linder, Jacob},
  journal = {Phys. Rev. B},
  volume = {113},
  issue = {5},
  pages = {054403},
  numpages = {7},
  year = {2026},
  month = {Feb},
  publisher = {American Physical Society},
  doi = {10.1103/zwz9-l7wf},
  url = {https://link.aps.org/doi/10.1103/zwz9-l7wf}
}

@article{Smejkal2022Landscape,
  author = {Smejkal, Libor and Sinova, Jairo and Jungwirth, Tomas},
  title = {Emerging Research Landscape of Altermagnetism},
  journal = {Phys. Rev. X},
  volume = {12},
  pages = {040501},
  year = {2022},
  doi = {10.1103/PhysRevX.12.040501}
}

@article{GonzalezHernandez2021,
  author = {Gonzalez-Hernandez, Rafael and Smejkal, Libor and Vyborny, Karel and Yahagi, Yuta and Sinova, Jairo and Jungwirth, Tomas and Zelezny, Jakub},
  title = {Efficient Electrical Spin Splitter Based on Nonrelativistic Collinear Antiferromagnetism},
  journal = {Phys. Rev. Lett.},
  volume = {126},
  pages = {127701},
  year = {2021},
  doi = {10.1103/PhysRevLett.126.127701}
}

@article{Smejkal2020SciAdv,
  author = {Smejkal, Libor and Gonzalez-Hernandez, Rafael and Jungwirth, Tomas and Sinova, Jairo},
  title = {Crystal Time-Reversal Symmetry Breaking and Spontaneous Hall Effect in Collinear Antiferromagnets},
  journal = {Sci. Adv.},
  volume = {6},
  pages = {eaaz8809},
  year = {2020},
  doi = {10.1126/sciadv.aaz8809}
}

@article{Bauer2012,
  author = {Bauer, G. E. W. and Saitoh, Eiji and van Wees, Bart J.},
  title = {Spin Caloritronics},
  journal = {Nat. Mater.},
  volume = {11},
  pages = {391--399},
  year = {2012},
  doi = {10.1038/nmat3301}
}

@article{Chumak2015,
  author = {Chumak, A. V. and Vasyuchka, V. I. and Serga, A. A. and Hillebrands, B.},
  title = {Magnon Spintronics},
  journal = {Nat. Phys.},
  volume = {11},
  pages = {453--461},
  year = {2015},
  doi = {10.1038/nphys3347}
}

@article{Smejkal2023ChiralMagnons,
  author = {Smejkal, Libor and Marmodoro, Alberto and Ahn, Kyo-Hoon and Gonzalez-Hernandez, Rafael and Turek, Ilja and Mankovsky, Sergiy and Ebert, Hubert and D'Souza, Sunil W. and Sipr, Ondrej and Sinova, Jairo and Jungwirth, Tomas},
  title = {Chiral Magnons in Altermagnetic {RuO2}},
  journal = {Phys. Rev. Lett.},
  volume = {131},
  pages = {256703},
  year = {2023},
  doi = {10.1103/PhysRevLett.131.256703}
}

@article{Cui2023EfficientSSE,
  title = {Efficient spin Seebeck and spin Nernst effects of magnons in altermagnets},
  author = {Cui, Qirui and Zeng, Bowen and Cui, Ping and Yu, Tao and Yang, Hongxin},
  journal = {Phys. Rev. B},
  volume = {108},
  issue = {18},
  pages = {L180401},
  numpages = {7},
  year = {2023},
  month = {Nov},
  publisher = {American Physical Society},
  doi = {10.1103/PhysRevB.108.L180401},
}

@article{Ni2010,
  author = {Ni, N. and Climent-Pascual, E. and Jia, S. and Huang, Q. and Cava, R. J.},
  title = {Physical Properties and Magnetic Structure of the Layered Oxyselenide {La2O3Mn2Se2}},
  journal = {Phys. Rev. B},
  volume = {82},
  pages = {214419},
  year = {2010},
  doi = {10.1103/PhysRevB.82.214419}
}

@article{Koo2012,
title = {Analysis of the magnetic structure of the manganese oxychalcogenides {R2Mn2Se2O (R=LaO, BaF)} by density functional calculations},
journal = {J. Magn. Magn. Mater.},
volume = {324},
number = {22},
pages = {3859-3862},
year = {2012},
issn = {0304-8853},
doi = {https://doi.org/10.1016/j.jmmm.2012.06.035},
url = {https://www.sciencedirect.com/science/article/pii/S0304885312005483},
author = {Hyun-Joo Koo and Myung-Hwan Whangbo},
}

@article{Lei2012,
  author = {Lei, Hechang and Bozin, Emil S. and Llobet, A. and Ivanovski, V. and Koteski, V. and Belosevic-Cavor, J. and Cekic, B. and Petrovic, C.},
  title = {Magnetism in {La2O3(Fe1-xMnx)2Se2} Tuned by {Fe/Mn} Ratio},
  journal = {Phys. Rev. B},
  volume = {86},
  pages = {125122},
  year = {2012},
  doi = {10.1103/PhysRevB.86.125122}
}

@article{Wei2024LMSO,
  title = {{La2O3Mn2Se2}: A correlated insulating layered d-wave altermagnet},
  author = {Wei, Chao-Chun and Li, Xiaoyin and Hatt, Sabrina and Huai, Xudong and Liu, Jue and Singh, Birender and Kim, Kyung-Mo and Fernandes, Rafael M. and Cardon, Paul and Zhao, Liuyan and Tran, Thao T. and Frandsen, Benjamin A. and Burch, Kenneth S. and Liu, Feng and Ji, Huiwen},
  journal = {Phys. Rev. Mater.},
  volume = {9},
  issue = {2},
  pages = {024402},
  numpages = {13},
  year = {2025},
  month = {Feb},
  publisher = {American Physical Society},
  doi = {10.1103/PhysRevMaterials.9.024402},
 }

@article{GarciaGassull2025LMSO,
  author  = {Laura Garcia-Gassull and Aleksandar Razpopov and P. Peter Stavropoulos and Igor I. Mazin and Roser Valent{\'\i}},
  title   = {Microscopic origin of the magnetic interactions and their experimental signatures in altermagnetic {La$_2$O$_3$Mn$_2$Se$_2$}},
  journal = {npj Spintronics},
  year    = {2026},
  volume  = {4},
  number  = {1},
  pages   = {9},
  doi     = {10.1038/s44306-025-00125-9},
}

@article{Li2025PressureLMSO,
  author  = {Yi-Kang Li and Ye Yang and Yuqing Zhang and Zhigang Gui and Xi-Kai Wen and Yan-Jun Li and Qingyuan Liu and Xian-Long Wang and Rui Wang and Jianjun Ying and Xianhui Chen},
  title   = {Unusual in-plane lattice collapse in layered {La$_2$O$_3$Mn$_2$Se$_2$} initiated by pressure-driven spin-crossover},
  journal = {Commun. Mater.},
  year    = {2025},
  volume  = {6},
  number  = {1},
  pages   = {245},
  doi     = {10.1038/s43246-025-00966-1},
}

@article{Asai2026LMSO,
  author = {Asai, Shinichiro and Hu, Junxi and Liu, Zheyuan and Itoh, Shinichi and Ueta, Daichi and Matsuda, Jin and Hatanaka, Tatsuto and Watanabe, Hikaru and Arita, Ryotaro and Masuda, Takatsugu},
  title = {Realization of a Two-Dimensional d-Wave Altermagnet in {La$_2$O$_3$Mn$_2$Se$_2$}},
  journal = {Phys. Rev. Mater.},
  volume = {10},
  pages = {L011401},
  year = {2026},
  doi = {10.1103/q863-3sfx}
}

@article{Kresse1993VASP,
  title = {Ab initio molecular dynamics for liquid metals},
  author = {Kresse, G. and Hafner, J.},
  journal = {Phys. Rev. B},
  volume = {47},
  issue = {1},
  pages = {558(R)--561(R)},
  numpages = {0},
  year = {1993},
  month = {Jan},
  publisher = {American Physical Society},
  doi = {10.1103/PhysRevB.47.558},
  url = {https://link.aps.org/doi/10.1103/PhysRevB.47.558}
}

@article{Kresse1996VASP,
  author = {Kresse, G. and Furthm{\"u}ller, J.},
  title = {Efficient iterative schemes for ab initio total-energy calculations using a plane-wave basis set},
  journal = {Phys. Rev. B},
  volume = {54},
  pages = {11169--11186},
  year = {1996},
  doi = {10.1103/PhysRevB.54.11169}
}

@article{Blochl1994PAW,
  author = {Bl{\"o}chl, P. E.},
  title = {Projector augmented-wave method},
  journal = {Phys. Rev. B},
  volume = {50},
  pages = {17953--17979},
  year = {1994},
  doi = {10.1103/PhysRevB.50.17953}
}

@article{Kresse1999PAW,
  author = {Kresse, G. and Joubert, D.},
  title = {From ultrasoft pseudopotentials to the projector augmented-wave method},
  journal = {Phys. Rev. B},
  volume = {59},
  pages = {1758--1775},
  year = {1999},
  doi = {10.1103/PhysRevB.59.1758}
}

@article{Perdew1996PBE,
  author = {Perdew, John P. and Burke, Kieron and Ernzerhof, Matthias},
  title = {Generalized Gradient Approximation Made Simple},
  journal = {Phys. Rev. Lett.},
  volume = {77},
  pages = {3865--3868},
  year = {1996},
  doi = {10.1103/PhysRevLett.77.3865}
}

@unpublished{SM,
  author     = {Rezaei, Nafise  and Qaiumzadeh, Alireza and Oganov, Artem R. and Alaei, Mojtaba},
  title     = {Supplementary Material for {``}Pressure-Tunable Electronic and Magnonic Transport in Altermagnet {La$_2$O$_3$Mn$_2$Se$_2$}{''}},
  note      = {Unpublished supplementary material},
  year      = {2026}
}

@article{Dudarev1998DFTU,
  author = {Dudarev, S. L. and Botton, G. A. and Savrasov, S. Y. and Humphreys, C. J. and Sutton, A. P.},
  title = {Electron-energy-loss spectra and the structural stability of nickel oxide: An {LSDA+U} study},
  journal = {Phys. Rev. B},
  volume = {57},
  pages = {1505--1509},
  year = {1998},
  doi = {10.1103/PhysRevB.57.1505}
}

@article{Alaei2024SUPERHEX,
  title = {Optimizing supercell structures for Heisenberg exchange interaction calculations},
  author = {Alaei, Mojtaba and Oganov, Artem R.},
  journal = {Phys. Rev. B},
  volume = {111},
  issue = {14},
  pages = {144419},
  numpages = {5},
  year = {2025},
  month = {Apr},
  publisher = {American Physical Society},
  doi = {10.1103/PhysRevB.111.144419},
}

@article{Rezaei2019ESpinS,
title = {{ESpinS}: A program for classical {Monte Carlo} simulations of spin systems},
journal = {	Comput. Mater. Sci.},
volume = {202},
pages = {110947},
year = {2022},
issn = {0927-0256},
doi = {https://doi.org/10.1016/j.commatsci.2021.110947},
author = {Nafise Rezaei and Mojtaba Alaei and Hadi Akbarzadeh},
}

@article{Hukushima1996ParallelTempering,
  author = {Hukushima, Koji and Nemoto, Koji},
  title = {Exchange {Monte Carlo} Method and Application to Spin Glass Simulations},
  journal = {	J. Phys. Soc. Jpn.},
  volume = {65},
  pages = {1604--1608},
  year = {1996},
  doi = {10.1143/JPSJ.65.1604}
}

@article{Mostofi2008Wannier90,
  author = {Mostofi, Arash A. and Yates, Jonathan R. and Lee, Young-Su and Souza, Ivo and Vanderbilt, David and Marzari, Nicola},
  title = {wannier90: A tool for obtaining maximally-localised Wannier functions},
  journal = {	Comput. Phys. Commun.},
  volume = {178},
  pages = {685--699},
  year = {2008},
  doi = {10.1016/j.cpc.2007.11.016}
}

@article{Marzari1997MLWF,
  author = {Marzari, Nicola and Vanderbilt, David},
  title = {Maximally localized generalized Wannier functions for composite energy bands},
  journal = {Phys. Rev. B},
  volume = {56},
  pages = {12847--12865},
  year = {1997},
  doi = {10.1103/PhysRevB.56.12847}
}

@article{Souza2001MLWF,
  author = {Souza, Ivo and Marzari, Nicola and Vanderbilt, David},
  title = {Maximally localized Wannier functions for entangled energy bands},
  journal = {Phys. Rev. B},
  volume = {65},
  pages = {035109},
  year = {2001},
  doi = {10.1103/PhysRevB.65.035109}
}

@article{Wang2006AHC,
  author = {Wang, Xinjie and Yates, Jonathan R. and Souza, Ivo and Vanderbilt, David},
  title = {Ab initio calculation of the anomalous Hall conductivity by Wannier interpolation},
  journal = {Phys. Rev. B},
  volume = {74},
  pages = {195118},
  year = {2006},
  doi = {10.1103/PhysRevB.74.195118}
}

@article{Toth2015SpinW,
  author = {Toth, Sandor and Lake, Bella},
  title = {Linear spin wave theory for single-Q incommensurate magnetic structures},
  journal = {J. Phys. Cond. Mat.},
  volume = {27},
  pages = {166002},
  year = {2015},
  doi = {10.1088/0953-8984/27/16/166002}
}

\end{document}